\documentstyle[12pt]{article}
\begin{document}
\title{\vspace*{-1.8cm}
Connecting nonleptonic and weak radiative
hyperon decays}
\author{
{P. \.Zenczykowski}$^*$\\
\\
{\em Dept. of Theoretical Physics},
{\em Institute of Nuclear Physics}\\
{\em Radzikowskiego 152,
31-342 Krak\'ow, Poland}\\
}
\maketitle
\begin{abstract}
Using the recent  measurement of the
$\Xi ^0 \to \Lambda \gamma $ asymmetry as an input,
we reanalyse nonleptonic  and weak radiative 
hyperon decays in a single symmetry-based framework.
In this framework
the old S:P problem of nonleptonic decays is 
automatically resolved when
the most important features of weak radiative decays
are taken into account as an input.
Experimental data require that symmetry between the
two types of hyperon decays 
be imposed at the level
of currents, not fields.
Previously established connections between hyperon decays
and nuclear parity violation
  imply that the
conflict, originally suggested by weak radiative decays, 
 has to surface somewhere. 
\end{abstract}
\\
$^*$ E-mail:
zenczyko@iblis.ifj.edu.pl\\
\phantom{$^*$} phone: (48-12)662-8273;
fax: (48-12)662-8458
\newpage

For a long time weak hyperon decays have been presenting us 
with a couple of puzzles (see \cite{DGH,LZ}). These have been in particular:
 the question of the S:P ratio in the nonleptonic hyperon 
decays (NLHD)
and the issue of a large negative asymmetry in the 
$\Sigma ^+ \to p \gamma $ weak radiative hyperon decay (WRHD), the latter
being indicative of either SU(3) breaking effects much larger than
expected, or of Hara's theorem being violated.
Violation of this theorem,
although forbidden on the basis of hadron-level arguments,
was also suggested by a couple of (technically correct) 
quark-level calculations (constituent quark model, CQM)
\cite{KR, Zen01,nonloc}. 
In the CQM calculations 
the constituent quarks in the intermediate states
between the action of weak interaction and the emission of a photon 
are essentially free.
Violation of Hara's theorem
followed also when NLHD and WRHD were connected
via the vector-meson dominance (VMD) 
approach \cite{Zen89}.

Some time ago it was pointed out \cite{LZ} that 
the status of Hara's theorem can be clarified through the measurement of
the $\Xi ^0 \to \Lambda \gamma $ decay asymmetry.
By yielding 
a large and negative value of $ -0.65 \pm 0.19 $
for this asymmetry, the
recent NA48 experiment \cite{Schmidt} has
decided very clearly in favour of the theorem.
The experimental number disagrees very strongly 
with the results of the CQM calculations 
\cite{VS,Zen01}, and with the VMD approach \cite{Zen01,Zen89,Zen91},
which both yield large positive value for this asymmetry.
Consequently, we are forced to conclude that 
{\em (a)} constituent quark
calculations do not provide us with a proper description of 
weak hyperon decays, 
and {\em (b)} the existing description of NLHD and/or 
the connection between NLHD and WRHD used in the VMD-based approach
of \cite{Zen89} 
do not correspond to physical reality.

The aim of this paper is to present a symmetry-based
 explanation of both the measured
S:P ratio in NLHD and the gross structure of the
observed pattern of asymmetries and branching
ratios in WRHD, an explanation which
maintains an intimate connection between NLHD and WRHD,
and yet does not lead to the CQM/VMD results
of \cite{KR,Zen89,VS,Zen91}.
Fragments of this explanation have been known for over twenty years
now \cite{LeY,Gav},
but they have never been presented in a way clearly highlighting
the underlying simple symmetry connection.

One generally expects that NLHD and WRHD 
should be related (see eg. \cite{Zen89,Paschos}),
and that this should hold both for the parity-conserving (p.c.) and
the parity-violating (p.v.) amplitudes.
Theoretical and phenomenological analyses 
of p.c. hyperon decay amplitudes have shown many
times that
 there are no basic problems here, only some numerological
 differences. 
 The p.c. NLHD amplitudes are sufficiently 
 well described by the pole model, while the
 p.c. WRHD amplitudes 
 may be estimated from NLHD through $SU(2)_W$ spin symmetry
 with the help of VMD (or in another effectively
 equivalent way), always
 leading to qualitatively similar results.
 Probably the most reliable phenomenological 
 evaluation of p.c. WRHD amplitudes along the symmetry lines was carried out
in ref.\cite{Zen91}.
The real problem, as troubles with Hara's theorem indicated, is with
the p.v. amplitudes.

In order to connect the asymmetries of NLHD with those of WRHD,
we need to be consistent when fixing the
relative signs between p.c. and p.v. amplitudes of both NLHD and WRHD.
Our conventions are the same as those in  
ref.\cite{Zen89} (Table IV, Eq. (5.2b)). 
For the purposes of this paper, it is sufficient to recall the 
following expressions for the p.c. amplitudes: 
\begin{eqnarray}
\label{eqar11}
B(\Xi ^0 \to \Lambda \pi ^0) &=& 
\frac{1}{2\sqrt{3}}\left[
\left(3\frac{f}{d}-1\right)\frac{F_A}{D_A}+3-\frac{f}{d}
\right]C\\
\label{eqar12}
B(\Xi ^0 \to \Lambda \gamma)& = & \frac{e}{g_{\rho}}\left[
-\frac{4}{3\sqrt{3}}C
\right]\\
\label{eqar13}
B(\Xi ^0 \to \Sigma ^0 \gamma )& = & \frac{e}{g_{\rho}}
\left[-\frac{4}{3}C\right]
\end{eqnarray}
where $g_{\rho}=5.0$, $F_A/D_A \approx 0.56$, and $f/d \approx -1.8$
or $-1.9$.
The  factor of $\frac{e}{g_{\rho}}$, reminiscent of VMD, follows from the
replacement of the strong coupling with the electromagnetic one, ie. its
appearance does not require VMD (but is consistent with it) \cite{Zen01}.
The $\Xi ^0 \to \Lambda \gamma$
and $\Xi ^0 \to \Sigma ^0 \gamma $ p.c. amplitudes are of the same sign.
Furthermore,
the ratio $B(\Xi ^0 \to \Lambda \gamma)/B(\Xi ^0 \to \Lambda \pi ^0)$
is around $-3e/g_{\rho}$, ie. negative.

For the p.v. amplitudes, with the
experiment forcing us to abandon the constituent quark description 
\cite{Zen01,Schmidt},
we also turn to hadron-level approaches. In the approach of ref.\cite{Zen89}, 
the WRHD p.v. amplitudes were calculated using symmetry from the NLHD p.v.
amplitudes
through the chain of connections: $\pi $  $\stackrel{SU(6)_W}{\to} $ 
$ \rho $ ($\omega, \phi $)
$\stackrel{VMD}{\to }$ $\gamma $.
The basic assumptions were $VMD$, $SU(6)_W$, and 
the assumption that the p.v. NLHD amplitudes 
are well described by the current-algebra (CA)
commutator term. 
In ref.\cite{KKW} it was proved
that the soft meson CA approach and the 
SU(6)-symmetric
quark-line diagram approach
are totally equivalent in a group-theoretical sense.
The perturbative QCD effects can yield only nonleading corrections to the
simple quark-diagram scheme. The confining 
effects require going from quark to hadron level of description,
in which quarks are treated as spin-flavour indices of effectively
local hadron fields. 
This is how the whole $SU(6)_W$ quark diagram scheme is understood.
Although in principle the nonperturbative effects could 
affect the simple quark diagram scheme, they are not expected
to do so:
the quark diagram scheme works well in many places,
somehow including all such effects
(see also the comment on p.338 of ref.\cite{DGH}). 
In conclusion, the quark-diagram approach to NLHD (ref. \cite{Zen89}) 
was completely consistent
with current algebra. When generalized to WRHD, this approach
predicted large positive asymmetry for the
$\Xi ^0 \to \Lambda \gamma $ decay \cite{Zen91}. As this prediction strongly
disagrees with the data, 
the following two questions emerge: \\
i) Is the input in the chain of connections (ie. the p.v. NLHD amplitudes) 
understood sufficiently well ?
 (The S:P puzzle indicates there may be a problem here.)\\
ii) Is the chain of connections itself correct (ie. are
$SU(6)_W$ and VMD applied in a proper way), and - if not - 
how to modify it?

The assumption of $SU(6)_W$ relates
the relative sizes of contributions to the p.v. amplitudes corresponding
to the diagrams shown in Fig.1, but only for contributions 
from a single class: 
either $(b1)$ or $(b2)$, etc. It does not relate $(b1)$ to $(b2)$ (or to
$(c1)$ or $(c2)$), and it does not connect NLHD with WRHD.
Fixing the relative sizes and signs of contributions from these four types
of diagrams,
both for NLHD and WRHD separately, as well as between NLHD and WRHD, 
requires additional assumptions that go beyond mere $SU(6)_W$.
In other words the $SU(6)_W$ quark diagram approach 
has to be properly augmented,
so that it indicates not only the contractions of quark indices, but also
the way in which various diagrammatic amplitudes are to be combined (ie. what
are their relative strengths and sizes).
For example, the soft meson term corresponds to a particular combination 
of diagrammatic amplitudes, as adopted in \cite{KKW}. However, CA admits
 a correction to this term (the correction being
 proportional to meson momentum), which corresponds to a different
 combination of diagrammatic amplitudes \cite{Zen99}.
Similarly, the constituent quark model fixes (in disagreement 
with experiment \cite{Zen01}) 
the relative
signs and sizes of all $(b)$-type NLHD and WRHD amplitudes.

The $SU(6)_W$ coefficients with which different amplitude types 
contribute to different decays were
calculated in the past (see \cite{Zen94}), and are
gathered in Table 1 for NLHD, and in Table 2 for WRHD. Table 2 shows
only the coefficients 
appropriate for $(b)$-type
 diagrams  as 
the smallness of the measured $\Xi ^- \to \Sigma ^- \gamma$ branching
ratio implies that the total single-quark
contribution (which involves $(c)$-type diagrams) is negligible.

\begin{table}[t]
\caption{$SU(6)_W$ coefficients for NLHD}
\label{table1}
\begin{center}
\begin{tabular}{cccccc}
\hline
& Decay $k$        & $b_1(k)$ & $b_2(k)$ & $c_1(k)$ & $c_2(k)$ \\
\hline
$\Sigma ^+_0$ & $\Sigma ^+ \to p \pi ^0$   & $0$  & $\frac{1}{2\sqrt{2}}$ 
& $-\frac{1}{6\sqrt{2}}$ & $0$ \\
$\Sigma ^-_-$ & $\Sigma ^- \to n \pi ^-$   & $0$  & $-\frac{1}{2} $ 
& $\frac{1}{6} $ & $0$\\
$\Lambda ^0_0$ & $\Lambda \to n \pi ^0 $   & $0$  & $\frac{1}{4\sqrt{3}}$ 
& $-\frac{1}{4\sqrt{3}}$ & $0$ \\
$\Xi ^0_0$     & $\Xi ^0 \to \Lambda ^0 \pi ^0$   & $0$  & 
$-\frac{1}{2\sqrt{3}}$ 
& $\frac{1}{4\sqrt{3}} $ & $0$\\
\hline
\end{tabular}
\end{center}
\end{table}

\begin{table}
\caption{$SU(6)_W$ coefficients for WRHD}
\label{table2}
\begin{center}
\begin{tabular}{ccc}
\hline
Decay $k$ &$b_1(k)$ & $b_2(k)$ \\
\hline
$\Sigma ^+ \to p \gamma $ &$-\frac{1}{3\sqrt{2}}$ & $-\frac{1}{3\sqrt{2}}$ \\
$\Lambda \to n \gamma   $ &$\frac{1}{6\sqrt{3}}$  & $\frac{1}{2\sqrt{3}}$  \\
$\Xi ^0 \to \Lambda \gamma $ & $0$ & $-\frac{1}{3\sqrt{3}}$ \\
$\Xi ^0 \to \Sigma ^0 \gamma$ & $\frac{1}{3}$ & $0$\\
\hline
\end{tabular}
\end{center}
\end{table}

Experimental p.v. NLHD amplitudes $A_{NL}(k)$ are 
phenomenologically very well described 
by
\begin{equation}
\label{expNLHD}
A_{NL}(k) = b_2(k) ~b_{NL} +c_1(k)~c_{NL}
\end{equation}
with
\begin{eqnarray}
\label{eqar21}
b_{NL}& = & -5\\
\label{eqar22}
c_{NL} &= & 12
\end{eqnarray}
(in units of $10^{-7}$),
which corresponds to the S-wave SU(3) parameters
\begin{eqnarray}
\label{fdS}
d_S&=&b_{NL}\\
f_S&=&-b_{NL}+\frac{2}{3}c_{NL}
\end{eqnarray}
satisfying
\begin{eqnarray}
\label{fdSratio}
\frac{f_S}{d_S}&=&-1+\frac{2}{3}\frac{c_{NL}}{b_{NL}}=-2.6
\end{eqnarray}

Experimental values of the
$\Xi ^0 \to \Lambda \gamma$  
and $\Xi ^0 \to \Sigma ^0 \gamma $ branching ratios
are well described  when, after factorizing the $b_i$
coefficients and replacing
the electromagnetic couplings in the WRHD amplitudes
with their strong counterparts,
the moduli of
the (rescaled) $SU(6)_W$ p.v. WRHD
amplitudes for diagrams (b1) and (b2) (Fig. 1) are both 
{\em numerically} (approximately)
equal to $|b_{NL}|$ :
\begin{equation}
|b_{WR}(b1)|=|b_{WR}(b2)|\approx |b_{NL}|
\end{equation}
(the two branching ratios in question do $\em not$
 depend on the signs of $b_{WR}(b1)$ and $b_{WR}(b2)$, 
 see \cite{LZ}).
 Incidentally, this agreement shows that the effects of SU(3)
 breaking in these two decays are not large, as discussed in
 refs. \cite{LZ,Zen94} (the dominant terms in the p.v. amplitudes
 are SU(3) symmetric).
Furthermore, with p.c. amplitudes of these decays being of equal sign
(Eqs.(\ref{eqar12},\ref{eqar13})), it follows from Table 2 that
equal signs of their experimental asymmetries require
that the total p.v. WRHD amplitudes
 $A_{WR}$
be proportional to the difference $b_1-b_2$:
\begin{equation}
\label{expWRHD}
A_{WR}(k) =  \frac{e}{g_{\rho}} (-b_1(k) + b_2(k)) ~b_{WR}
\end{equation}
with factor $\frac{e}{g_{\rho}}$ accounting for the replacement of the strong
coupling with the electromagnetic one, just as in p.c. amplitudes
(Eqs.(\ref{eqar12},\ref{eqar13})).
Finally, with p.c. amplitudes of $\Xi ^0 \to \Lambda \pi ^0$ and
$\Xi ^0 \to \Lambda \gamma $ decays being of opposite signs 
(Eqs (\ref{eqar11},\ref{eqar12})), from the
equal signs of their
experimental asymmetries
it follows (using Tables 1 and 2 and Eqs (\ref{eqar21},\ref{eqar22})) that
\begin{equation}
\label{bWRbNL}
b_{WR}\approx -b_{NL}. 
\end{equation}

The form of Eq.(\ref{expWRHD}) ensures that Hara's theorem holds
for exact SU(3) since the SU(3)-symmetric $b_i$ terms cancel there (Table 2).  
As proposed in ref.\cite{Gav}, the
large value of the experimental $\Sigma ^+ \to p \gamma $
asymmetry is presumably due to a substantial SU(3) breaking effect,
expected to be of the 
order of $\delta s/\delta \omega$, with the SU(3) breaking mass difference
$\delta s \approx
m_s-m_d \approx 190 ~MeV$, and $\delta \omega \approx 570 ~MeV$ being
the energy difference between the first excited $1/2^-$ state and the 
ground state.

We now proceed to
the question of the theoretical connection between NLHD and WRHD, ie.
between Eq.(\ref{expNLHD}) 
and Eq.(\ref{expWRHD}). 
Current algebra expresses the p.v. NLHD amplitudes $A_{NL}(k)$ 
in terms of the
contribution $C(k)$ from the CA commutator and the correction 
$q^{\mu}R_{\mu}(k)$
proportional
to the momentum of the emitted pion:
\begin{equation}
\label{CA}
A_{NL}(k) = C(k) + q^{\mu}R_{\mu}(k)
\end{equation}
Ref.\cite{Zen01} proves
that the $b$-diagram-dependent part of $C(k)$ is proportional
to $b_1(k)+b_2(k)$. It may be shown 
using the derivative form of the strong coupling of $\pi$ to baryons
(see \cite{Zen01,Zen99}),
that
the $b$-diagram-dependent part of the
$q^{\mu }R_{\mu}(k)$ term is proportional to $-b_1(k)+b_2(k)$.
Namely, by CP invariance and hermiticity,
in the parity-violating CP-conserving couplings of $CP=-1$
neutral pseudoscalar mesons to baryons
\begin{equation}
\label{symmetries}
g^{(0)}_{fi}\bar{u}_fu_iP^0+
g^{(1)}_{fi}q^{\mu}\bar{u}_f\gamma _{\mu}u_iP^0,
\end{equation}
in the convention of
\cite{Zen99},
the coefficients $g^{(n)}$ are imaginary,
with $g^{(0)}_{fi}$ ($g^{(1)}_{fi}$) 
antisymmetric (symmetric) under
$i \leftrightarrow f$ interchange.
When translated into the language of $b_i$ coefficients,
this leads to $b_1 + b_2$ and $-b_1+b_2$ structures for the
$q$-independent and $q$-dependent terms respectively \cite{Zen01,Zen99}.
$SU(2)_W$ spin symmetry relates contributions to NLHD and WRHD from terms
proportional to $-b_1(k)+b_2(k)$.
Since experimentally
the total contribution of all single-quark diagrams 
(including the $(c)$-type ones) is negligible in WRHD, 
for NLHD we expect from symmetry with WRHD
that a substantial
contribution from $(c)$-type diagrams 
arises from the commutator term only.

Thus, the soft-meson CA approach of \cite{KKW} is augmented
with a correction term, due to a non-zero value of pion momentum
and estimated from WRHD by symmetry:
\begin{equation}
\label{fullNLHD}
A_{NL}(k)= (b_1(k)+b_2(k))~b_{com}+(c_1(k)+c_2(k))~c_{com}
+(-b_1(k)+b_2(k))~b_{WR}
\end{equation}
where the first two terms describe the commutator $C(k)$, and
the symmetry between NLHD and WRHD
is used for the $-b_1+b_2$ term.

Consequently,
\begin{eqnarray}
b_{NL} & = & b_{com} + b_{WR}\\
c_{NL} & = & c_{com}
\end{eqnarray}
Note that for NLHD all $b_1(k)$ are zero and, consequently,
without knowing $b_{WR}$
we cannot extract $b_{com}$ directly from the data. Clearly,
we cannot have $b_{com}=0$ as suggested by the
constituent quark model calculations
combined with Hara's theorem 
\cite{Zen01}, or else we would have $b_{NL}=b_{WR}$, and
the $\Xi ^0 \to (\Lambda ,\Sigma ^0 )\gamma $ 
asymmetries would be predicted as positive, in disagreement
with experiment.

Using Eq.(\ref{bWRbNL}) we obtain
\begin{equation}
b_{com}\approx 2b_{NL}
\end{equation}
and 
for the commutator we have
\begin{eqnarray}
d_{com}&=&2b_{NL}\\
f_{com}&=&-2b_{NL}+\frac{2}{3}c_{NL}
\end{eqnarray}
Consequently
\begin{eqnarray}
d_{com}/d_S& = & 2\\
f_{com}/f_S& \approx & 1.4\\
\frac{f_{com}}{d_{com}}&=&-1+\frac{1}{3}\frac{c_{NL}}{b_{NL}} = -1.8.
\end{eqnarray}

Thus, the values of $d_{com}$ and $f_{com}$ 
(extracted from the S-wave amplitudes) agree with
the values of 
SU(3) parameters $d_P$ and $f_P$ needed to describe the P-wave amplitudes.
The resolution of the S:P problem and the description of WRHD
are interconnected.
The mechanism by which the $S$-wave amplitudes are reduced
from their commutator values is closely related to the
explanation proposed in ref.\cite{LeY}. In ref. \cite{LeY}
the downward correction is due to the $(70,1^-)$ intermediate
states. In our approach
explicit intermediate states are not used.
However, the symmetry properties of the correction term in
ref.\cite{LeY} and in this paper are identical in the symmetry limit.
The difference is that in this paper, 
instead of estimating the overall size of the correction in a model
as the authors of ref.\cite{LeY} do, we extract both its size and sign
 from WRHD.

There still remains a question how to understand the absence in WRHD 
of a term
proportional to
$b_1+b_2$ (ie. the
analogue of the CA commutator term as obtained in the constituent quark model
calculations).
We observe that if 
in the
presence of weak (p.v.) perturbation ${\cal L}^{p.v.}$
the symmetry is imposed between axial and 
vector currents $J^{\mu }_A$ and $J^{\mu }_V$,
the resulting couplings to photons and pions are 
obtained from
\begin{eqnarray}
\label{JV}
A_{\mu} T(J^{\mu }_V (x) {{\cal L }^{p.v.}(0)})&&\\
\label{JA}
\partial_{\mu} T(J^{\mu }_A (x){\cal L }^{p.v.}(0)) &=
& T(\partial _{\mu} J^{\mu }_A (x){{\cal L }^{p.v.}(0)}) + {\rm commutator}
\end{eqnarray}
 with the pion field appearing via PCAC in the first term on the r.h.s.
 of Eq.(\ref{JA}).
As shown in Eqs.(\ref{JV},\ref{JA}), the symmetry is not between the
pion field $\pi \propto \partial _{\mu} J^{\mu }_A$ and the photon field
$A_{\mu }$ (or the vector-meson field through VMD current-field identity
$J^{\mu }_V \propto V^{\mu }$)
but rather between the currents $J_V,J_A$ appearing on the l.h.s.
This bring us back to the original Gell-Mann's paper \cite{GM1964}.

This identification of symmetry necessary for a successful
joint description of nonleptonic and radiative weak hyperon decays
leads to problems elsewhere, however.
Namely, our present understanding of nuclear parity violation (cf. 
ref.\cite{DDH})
is based on symmetry of weak couplings between the {\em fields}
of pseudoscalar and vector mesons (and not on symmetry between
the axial and vector {\em currents}).
According to refs \cite{DDH,Despl} the explanation of data on nuclear parity
violation requires the dominance of the weak
rho-nucleon coupling of the form $\bar{u}_N\gamma _{\mu} \gamma _5 u_N
{\rho }^{\mu}$. Via the current-field identity (VMD) this leads 
to photon-nucleon coupling $\bar{u}_N\gamma _{\mu} \gamma _5 u_N
{A}^{\mu}$
and the violation of Hara's theorem in weak radiative hyperon
decays \cite{LZ,Zen89}.
Since the negative asymmetry of the $\Xi^0 \to \Lambda \gamma$ decay 
means that
Hara's theorem is satisfied, it follows
that either 
the current-field identity is not universal
or our present understanding of
nuclear parity violation
(ie. \cite{DDH})  
is not fully correct.

In conclusion:\\
1) The simple constituent quark model 
may produce unphysical results in higher order calculations
if free constituent quarks are used in intermediate states.
Consequently, it is an
 idealization that goes too far, and
 should be used and interpreted with care.
 The constituent quark model should better be regarded 
 as a method of evaluating symmetry properties of simplest hadronic
 couplings and transitions.\\
2) The connection between nonleptonic and weak radiative hyperon decays
should be formulated 
at the level of hadronic currents $J_A$, $J_V$ 
(and not at the level of
fields $\pi$,
$\rho$, $\gamma $) in agreement with Gell-Mann's
paper \cite{GM1964}.\\ 
3) The sizes and signs of the p.v. WRHD amplitudes are correlated
with those of the {\em correction} to the commutator term
in NLHD. When WRHD data are used to estimate this  correction,  
the old S:P problem in NLHD is resolved.
The explanation of the large asymmetry in $\Sigma ^+ \to p \gamma $
presumably requires 
more detailed SU(3)-breaking considerations (eg. \cite{Gav}).\\
4) The current-field identity suggests that vector mesons do not couple
to baryons through the $\bar{u}\gamma _{\mu }\gamma _5 u V^{\mu}$ term.
This is in conflict with our understanding of nuclear parity
violation \cite{Despl}. Thus, either this understanding 
 is not fully correct, or current-field identity 
is not universal.

{\bf Acknowledgements}\\
This work was supported in part by the
Polish State Committee for Scientific Research 
grant 5 P03B 050 21.

\vfill
\newpage
FIGURE CAPTION

Fig.1 $SU(6)_W$ diagrams for weak hyperon decays
\vfill

\begin{thebibliography}{99}
\bibitem{nonloc} In a gauge-invariant framework, 
the violation of Hara's theorem
requires that essential nonlocality be present in the parity-violating 
photon coupling to baryons (see \cite{Zen01}).
\end{thebibliography}
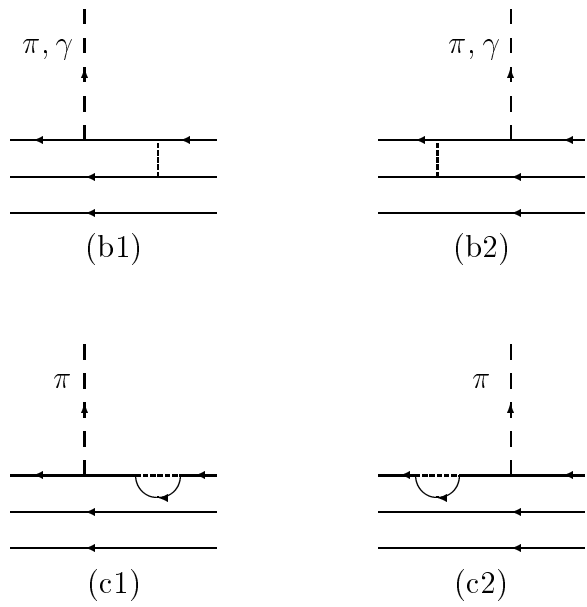
\end{document}